\documentclass[10pt,twocolumn]{article}

\usepackage[margin=0.75in,top=1in,bottom=1in]{geometry}
\usepackage{fontspec}
\defaultfontfeatures{Ligatures=TeX}
\newcommand{\headingfont}{\rmfamily}
\usepackage{amsmath}
\usepackage{amssymb}
\usepackage{graphicx}
\usepackage{booktabs}
\usepackage{hyperref}
\usepackage{url}
\usepackage[numbers,sort&compress]{natbib}
\usepackage{xcolor}
\usepackage[protrusion=true,expansion=false]{microtype}
\usepackage{caption}
\usepackage{subcaption}
\usepackage{tikz}
\usetikzlibrary{shapes.geometric, arrows.meta, positioning, fit, backgrounds}
\usepackage{float}

\hypersetup{
    colorlinks=true,
    linkcolor=blue,
    citecolor=blue,
    urlcolor=blue
}

\makeatletter
\renewcommand\section{\@startsection{section}{1}{\z@}%
  {-2.3ex \@plus -1ex \@minus -.2ex}%
  {1ex \@plus .2ex}%
  {\headingfont\Large\bfseries}}
\renewcommand\subsection{\@startsection{subsection}{2}{\z@}%
  {-2ex\@plus -1ex \@minus -.2ex}%
  {0.75ex \@plus .2ex}%
  {\headingfont\large\bfseries}}
\renewcommand\subsubsection{\@startsection{subsubsection}{3}{\z@}%
  {-1.8ex\@plus -1ex \@minus -.2ex}%
  {0.5ex \@plus .2ex}%
  {\headingfont\normalsize\bfseries}}
\makeatother

\title{
  {\headingfont\Large\bfseries Benchmarking Sensor Robustness in Plasma Diagnostic Models:}\\
  {\headingfont\Large\bfseries A Systematic Evaluation on TokaMark}
}

\author{
    Neerav Gupta \\
    \small Independent Researcher \\
    \small \texttt{neerav.dgupta@gmail.com}
}

\date{}

\begin{document}
\maketitle

\begin{abstract}
Plasma diagnostic models for tokamak fusion devices are almost universally
evaluated on clean, complete sensor data. In practice, fusion diagnostics
fail regularly: acquisition systems start late, individual sensors die, and
signal dropouts cluster precisely when a plasma disruption is approaching.
We present the first systematic robustness benchmark for plasma diagnostic
ML using the TokaMark dataset of 11,573 MAST shots~\cite{rousseau2026tokamark},
evaluating XGBoost, LSTM, Transformer, and the TokaMark CNN baseline across
six physically-grounded failure scenarios and three imputation strategies.
We introduce the Robustness Score (RS) for standardized cross-architecture
comparison. Our central finding is that disruption-proximate sensor
failure---corruption injected in the final window timesteps---collapses
sequence model performance (LSTM $+212\%$ NRMSE) while a statistical
feature model remains comparatively stable (XGBoost $+37\%$). Forward-fill
imputation eliminates nearly all degradation from random dropout for
sequence models (LSTM $+57\% \to {\sim}0\%$), but offers little help
when the end of the window is corrupted. Shot-level alarm evaluation using
ground-truth disruption timestamps reveals that LSTM alarm detection
collapses to TPR\,=\,0.00 under proximate sensor failure, while mean-fill
imputation recovers it to TPR\,=\,1.00, a reversal of the pattern observed
in NRMSE. Plasma current emerges as the single most critical diagnostic
across all architectures ($+73\%$ to $+140\%$ upon removal). Code, data,
and trained checkpoints are available at
\url{https://github.com/Neerav-Gupta/tokamark-robustness}.
\end{abstract}

\section{Introduction}

\subsection{Nuclear Fusion and Tokamaks}

Nuclear fusion has long been considered one of the most compelling
paths toward clean, abundant energy. Where fission splits heavy atoms
and leaves behind radioactive byproducts, fusion joins light hydrogen
isotopes (deuterium and tritium) and releases substantially more
energy per reaction, with helium as the main product~\cite{freidberg2007}.
The fuel supply is effectively inexhaustible: deuterium is extracted
from seawater, and tritium can be bred from lithium. A working fusion
power plant would emit no carbon, produce no long-lived radioactive
waste, and carry no risk of a runaway reaction~\cite{wesson2004}.

The central difficulty is confining the fuel long enough for fusion to
occur. The reaction requires temperatures above 100 million degrees
Celsius, hotter than the core of the Sun, at which hydrogen exists
as a plasma that would instantly destroy any material it contacted.
Magnetic confinement fusion (MCF) handles this by using carefully shaped
magnetic fields to keep the plasma suspended away from reactor
walls~\cite{freidberg2007}. Among MCF concepts, the tokamak has proven
the most successful, and ITER, now under construction in southern
France at an estimated cost of \$22 billion, is the international
experiment intended to demonstrate net energy gain from fusion for the
first time~\cite{iter1999}.

\subsection{Disruptions: The Critical Safety Challenge}

A plasma disruption is a sudden, uncontrolled loss of plasma confinement:
within milliseconds, the thermal and magnetic energy stored in the plasma
dumps into the reactor walls~\cite{devries2011}. De~Vries et
al.~\cite{devries2011} catalogued disruptions at JET, the Joint European
Torus and the world's largest operating tokamak, finding they occur in
roughly 3--4\% of routine plasma shots, with higher rates under demanding
operating conditions. At the scale of ITER or a commercial plant, each
disruption inflicts heat loads, electromagnetic impulses, and runaway
electron beams capable of permanently damaging the first wall, magnets,
and structural components~\cite{iter1999}.

The economic stakes are considerable. Maris et al.~\cite{maris2024}
showed that even moderate disruption rates push up the levelized cost
of electricity enough to threaten the commercial viability of fusion
energy. The practical response is early disruption prediction: if a
machine's control system receives even 50--100 milliseconds of warning,
it can gracefully terminate the plasma before damage
occurs~\cite{katesharbeck2019}. Reliable, real-time prediction is
therefore not merely a performance goal; it is a safety requirement
for any reactor-class device.

\subsection{AI for Plasma Prediction}

Plasma physics is complex and the diagnostic environment is demanding,
which together create a strong case for data-driven methods. Fusion
diagnostics cover magnetics, optical emission, X-ray cameras, microwave
interferometry, and more, spanning sampling rates from 200 Hz to 500 kHz
and producing inherently multi-modal, asynchronous, and noisy
data~\cite{morris2002}. Machine learning models can fuse these
heterogeneous streams, respond in real time, and capture nonlinear
dynamics without requiring explicit physical parameterization~\cite{churchill2025}.

Progress has been rapid. Kates-Harbeck et al.~\cite{katesharbeck2019}
showed that deep recurrent networks trained on raw diagnostic time series
could predict disruptions across JET and DIII-D with strong performance.
Vega et al.~\cite{vega2022} took this further, deploying an AI-based
disruption prediction system at JET that matched or outperformed existing
physics-based predictors. Churchill et al.~\cite{churchill2020} demonstrated
that convolutional networks operating directly on high-resolution
diagnostics could reliably identify pre-disruption states on DIII-D.
More recently, foundation model approaches have been proposed as a
path toward general-purpose plasma representations: Churchill~\cite{churchill2025}
laid out the theoretical case, and TokaMind~\cite{boschi2026tokamind}
delivered the first open-source multi-modal transformer trained on MAST
data and benchmarked on TokaMark tasks.

\subsection{The Missing Robustness Problem}

Despite this progress, almost every published evaluation rests on the
same unstated assumption: that sensor data arriving at inference time
is complete and clean. In practice this assumption regularly fails.
Acquisition systems start late, producing front-loaded gaps that cover
10--16\% of signal history. Individual sensors fail mid-shot due to
heat damage and electromagnetic interference. And these failures are
not uniformly distributed in time; they tend to occur precisely when
the plasma is most unstable and prediction is most critical. A
disruption produces intense heat loads and electromagnetic transients
that actively degrade sensor performance in the moments leading up to
the event~\cite{rousseau2026tokamark, jackson2025ieee}.

TokaMark~\cite{rousseau2026tokamark} acknowledges this directly,
identifying robustness to incomplete state information as one of its
four core benchmark dimensions and noting that signals ``may be absent
for a given shot due to hardware issues'' or ``contain missing time
segments due to limited acquisition windows or diagnostic failures.''
Yet no evaluation of this dimension appears anywhere in the TokaMark
paper, and no prior work has systematically studied how plasma
diagnostic models degrade under controlled sensor failure.

This paper addresses that gap directly. Our contributions are:

\begin{enumerate}
    \renewcommand{\labelenumi}{(\arabic{enumi})}
    \item \textbf{Natural missingness characterization.} We stream
    300 MAST shots from FAIR-MAST and characterize the real NaN
    structure of each diagnostic category, finding that failures
    are front-loaded and category-correlated, directly informing
    our scenario design.

    \item \textbf{Six physically-grounded failure scenarios.} We
    implement random dropout, channel ablation, temporal gaps (front,
    random, pre-event), correlated diagnostic group failure, and
    disruption-proximate failure. The last scenario, to our knowledge,
    has not been evaluated in any prior work.

    \item \textbf{Cross-architecture robustness benchmark.} We evaluate
    XGBoost, LSTM, Transformer, and the TokaMark CNN baseline on
    Task~4-4 under all six scenarios, and introduce the Robustness
    Score (RS) as a standardized metric we encourage the community
    to adopt.

    \item \textbf{Shot-level alarm evaluation.} Using ground-truth
    disruption timestamps (\texttt{t\_cut}) from FAIR-MAST, we compute
    true positive rate (TPR) and mean warning time (MWT) for each model
    under clean and corrupted conditions.

    \item \textbf{Mitigation strategy evaluation.} We evaluate zero-fill,
    mean-fill, and forward-fill imputation across all scenarios for both
    NRMSE and shot-level alarm metrics, revealing that the optimal
    strategy differs between the two evaluation regimes.
\end{enumerate}

\section{Related Work}

\paragraph{TokaMark and FAIR-MAST.}
Rousseau et al.~\cite{rousseau2026tokamark} introduced TokaMark as the
first large-scale, openly available benchmark for AI models on real
fusion data, defining 14 tasks across four groups over an 80/10/10
shot-level split drawn from the FAIR-MAST
dataset~\cite{jackson2024softwarex, jackson2025ieee}. Their CNN baseline
reaches NRMSE 0.429 on Task~4-4 using the full 9,270-shot training set.
TokaMark lists robustness to incomplete state information as one of its
core evaluation dimensions but stops short of actually evaluating it.
The present work takes up that thread.

\paragraph{Disruption prediction.}
Machine learning for disruption prediction has attracted sustained
effort over the past decade. Kates-Harbeck et al.~\cite{katesharbeck2019}
established that cross-device deep learning was feasible on JET and
DIII-D data. Vega et al.~\cite{vega2022} brought an AI predictor into
real-time operation at JET. Churchill et al.~\cite{churchill2020} explored
raw high-resolution diagnostic inputs on DIII-D. Aymerich et
al.~\cite{aymerich2022} incorporated spatiotemporal profile structure
at JET, and Guo et al.~\cite{guo2021} applied similar approaches to EAST.
Every one of these studies assumes complete, clean sensor availability
at inference time, an assumption our work directly interrogates.

\paragraph{Disruption alarm systems.}
Rea et al.~\cite{rea2019} deployed a real-time ML disruption predictor
on DIII-D, establishing the standard per-shot alarm metric framework
(true positive rate, false alarm rate, warning time) used in the field.
Saperstein et al.~\cite{saperstein2025} designed an off-normal warning
system for SPARC. Ai et al.~\cite{ai2025} developed adaptive anomaly
detection for disruption prediction from first discharge. Spangher et
al.~\cite{spangher2025} introduced DisruptionBench as a systematic
evaluation framework. All of these works assume complete sensor
availability; our work evaluates what happens when that assumption fails.

\paragraph{Autoregressive plasma state forecasting.}
Several recent works use autoregressive sequence models for full-shot
plasma state prediction in reactor control contexts. Char et
al.~\cite{char2024} demonstrated deep recurrent networks for full-shot
prediction on DIII-D. Wang et al.~\cite{wang2025popsim} introduced the
POPSIM framework for autoregressive plasma state forecasting. Jalalvand
et al.~\cite{jalalvand2025} applied multimodal super-resolution to
discover hidden physics in fusion plasmas. These works share our
sequence-based modeling spirit but focus on clean-data prediction rather
than robustness under sensor failure.

\paragraph{Missing data in tokamak ML.}
Two prior works come closest to our setting.
Chandrasekaran et al.~\cite{golem2021} tested CatBoost on 187 GOLEM
shots with missing values, tracing most degradation to interferometer
outages. Yang et al.~\cite{yang2025hl3} tackled the related problem of
channels that are entirely absent when a model is deployed on a new
device, proposing learned pseudo-data placeholders for the HL-3
disruption predictor. Both studies are narrower than ours in scope:
we evaluate four model architectures across six distinct failure
modes at multiple severities, with three imputation strategies, and
introduce the first standardized robustness metric for this class of
problem.

\paragraph{Data augmentation for cross-device generalization.}
Chayapathy et al.~\cite{viewmakers2024} developed time-series viewmakers,
learned augmentations combining jittering, slicing, and temporal
warping, to improve disruption prediction robustness when transferring
between tokamak devices via the DisruptionBench framework. That work
is concerned with distribution shift across machines, not sensor
failure during a single inference, and is complementary to ours.

\paragraph{Foundation models for fusion.}
TokaMind~\cite{boschi2026tokamind} is the first open-source multi-modal
transformer foundation model for tokamak plasma dynamics, trained on
MAST data. Churchill~\cite{churchill2025} made the broader case for
foundation model approaches in fusion. Both acknowledge missing-channel
handling as a real deployment challenge but do not quantify the
resulting degradation, which is what we provide.

\paragraph{XGBoost for scientific time series.}
Chen and Guestrin~\cite{chen2016xgboost} designed XGBoost with a
sparsity-aware split-finding algorithm that handles missing feature
values natively without any imputation step. This makes it a natural
comparison point for studying how architectural inductive bias shapes
robustness in sequence-based deep learning models.

\section{Background}

\subsection{The MAST Tokamak and FAIR-MAST}

MAST was a spherical tokamak operated at Culham, UK by the UK Atomic
Energy Authority from 1999 to 2013~\cite{sykes2001, meyer2009}. Over
its lifetime it produced more than 30,000 plasma shots, each lasting
roughly 2--3 seconds. Its compact spherical geometry distinguished it
from conventional tokamaks and made it a particularly useful testbed
for studying plasma dynamics and disruption physics relevant to future
spherical power plant concepts~\cite{counsell2005}.

FAIR-MAST~\cite{jackson2024softwarex, jackson2025ieee} makes diagnostic
data from 11,573 shots across MAST's final five experimental campaigns
openly available in Zarr format via a public S3 endpoint. Each shot
includes a ground-truth disruption timestamp (\texttt{t\_cut}), the
time at which plasma current quench occurs. TokaMark~\cite{rousseau2026tokamark}
draws on this data, selecting 39 signals and defining 14 prediction
tasks with an 80/10/10 shot-level split: 9,270 training shots, 1,157
validation shots, and 1,146 test shots.

\subsection{Task 4-4: Plasma Current Quench Prediction}

Among TokaMark's 14 tasks, we focus on Task~4-4: given 150 milliseconds
of diagnostic history, predict the plasma current 100 milliseconds into
the future. This is a non-Markovian autoregressive forecasting task;
the full input history is needed, not just the current
state~\cite{rousseau2026tokamark}. It is also the most directly
safety-relevant task in the benchmark, since plasma current quench is
one of the clearest precursors of a full disruption, and any prediction
system operating in a real device must handle it reliably.

The task takes 14 diagnostic input signals and 4 actuator signals as
input, listed in Table~\ref{tab:signals}. These span the full range of
MAST diagnostic categories: flux loops and pickup coils in magnetics,
interferometer and D-alpha in kinetics, soft X-ray in radiatives, and
active coil currents, plus the four actuator channels encoding
reference trajectories and fueling.

\begin{table}[H]
\centering
\caption{Input signals used in Task~4-4, with physical category and
channel count in the $(600, 18)$ time-series representation.}
\label{tab:signals}
\footnotesize
\begin{tabular}{lll}
\toprule
Index & Signal & Category \\
\midrule
0  & interferometer-n\_e\_line         & Kinetics \\
1  & magnetics-ccbv\_field             & Magnetics \\
2  & magnetics-obr\_field              & Magnetics \\
3  & magnetics-obv\_field              & Magnetics \\
4  & magnetics-omv\_voltage            & Magnetics \\
5  & magnetics-cc\_field               & Magnetics \\
6  & magnetics-saddle\_voltage         & Magnetics \\
7  & magnetics-flux\_loop\_flux        & Magnetics \\
8  & pf\_active-coil\_current          & Active coils \\
9  & pf\_active-solenoid\_current      & Active coils \\
10 & soft\_x\_rays-cam\_lower          & Radiatives \\
11 & soft\_x\_rays-cam\_upper          & Radiatives \\
12 & spectrometer-dalpha\_voltage      & Kinetics \\
13 & summary-ip                        & Plasma current \\
14 & gas\_injection-total\_injected    & Actuator \\
15 & pulse\_schedule-i\_plasma         & Actuator \\
16 & pulse\_schedule-n\_e\_line        & Actuator \\
17 & summary-power\_nbi                & Actuator \\
\bottomrule
\end{tabular}
\end{table}

\subsection{Natural Missingness in MAST}
\label{sec:missingness}

To ground our corruption scenarios in real failure patterns rather than
arbitrary assumptions, we streamed 300 shots from FAIR-MAST and measured
the NaN structure of each diagnostic category.
Table~\ref{tab:missingness} summarizes the results. Two findings shape
everything that follows.

\textbf{Category-stratified missingness.} Plasma current and coil
currents are present in every shot (0\% NaN across all 300), reflecting
the engineering priority placed on these signals for real-time control.
Kinetics diagnostics (interferometer and D-alpha) average 14--15\% NaN
per shot, caused by acquisition delays that consistently produce a
front-loaded gap at the start of each shot.

\textbf{Perfectly correlated failures.} The NaN percentages for
interferometer and D-alpha are identical to four decimal places in every
sampled shot, a correlation of $r = 1.000$. The two diagnostics share
a single acquisition trigger: when that trigger fires late, both signals
are simultaneously absent. This is not a statistical pattern but a
hardware reality, and it directly motivates the correlated group failure
scenario in Section~\ref{sec:scenarios}.

\begin{table}[H]
\centering
\caption{Natural NaN characterization across 300 sampled MAST shots.
Front block = single contiguous gap at start of shot.}
\label{tab:missingness}
\footnotesize
\resizebox{\columnwidth}{!}{%
\begin{tabular}{lrrrl}
\toprule
Signal category & Mean & Max & Missing & Structure \\
                & NaN\% & NaN\% & shots & \\
\midrule
Magnetics (ip, coils) & 0.0  & 0.0  & 0/300  & Always present \\
Flux loops            & 3.7  & 33.3 & 0/300  & Variable \\
Soft X-ray            & 1.9  & 34.1 & 15/300 & Variable \\
D-alpha spectrometer  & 14.4 & 42.5 & 7/300  & Front block \\
Interferometer        & 15.0 & 42.5 & 23/300 & Front block \\
\bottomrule
\end{tabular}%
}
\end{table}

\section{Methodology}

\subsection{Model Architectures}

We evaluate four architectures representing fundamentally different
inductive biases for multivariate time-series prediction.

\paragraph{XGBoost~\cite{chen2016xgboost}.}
Rather than operating on raw time series, we first compress each window
into 142 summary statistics: 9 per input signal (mean, std, min, max,
median, first value, last value, linear slope, and zero fraction across
14 input signals) and 4 per actuator signal (mean, std, first, last
across 4 actuators). This discards all temporal structure but compactly
captures the distributional properties of each signal. The XGBoost
regressor uses 500 estimators, learning rate 0.05, max depth 6,
subsample 0.8, colsample 0.8, and GPU histogram-based tree building
with early stopping after 20 rounds of no improvement on validation
NRMSE. Total split capacity: $\sim$500K decision tree nodes.

\paragraph{LSTM.}
Each diagnostic signal is spatially averaged to a scalar time series
and resampled to a fixed 600 timesteps by linear interpolation,
yielding a $(600, 18)$ tensor input. A 2-layer LSTM (hidden size 128,
dropout 0.2) processes this sequence and the final hidden state passes
through a two-layer MLP ($128 \to 64 \to 1$) to produce the
prediction. We train with Adam at learning rate $10^{-3}$,
ReduceLROnPlateau scheduling (patience 5, factor 0.5), gradient
clipping at 1.0, and early stopping with patience 15. Total parameters:
$\sim$430K.

\paragraph{Transformer.}
The same $(600, 18)$ tensor is linearly projected into a 128-dimensional
space, enhanced with sinusoidal positional encodings, and passed through
a 3-layer Transformer encoder (4 attention heads, feedforward dimension
256, dropout 0.1). A global average pool over the time axis feeds the
same MLP head ($128 \to 64 \to 1$). Training uses Adam with cosine
annealing ($T_\mathrm{max}=100$, $\eta_\mathrm{min}=10^{-5}$), gradient
clipping at 1.0, and early stopping with patience 15. Total parameters:
$\sim$310K.

\paragraph{CNN (TokaMark baseline).}
We implement the multi-branch convolutional encoder described in
Rousseau et al.~\cite{rousseau2026tokamark}, providing a direct
comparison to the TokaMark-reported baseline. Each of the 18 input
channels is processed by a dedicated 1D convolutional encoder ($N=3$
layers, kernel $K=3$, stride $s=3$, padding $p=1$, latent dimension
$D=16$). The resulting per-channel latent vectors are concatenated and
passed through a shared MLP backbone ($64 \to 32 \to 16 \to 1$).
Training uses Adam at learning rate $10^{-3}$, ReduceLROnPlateau
scheduling (patience 5, factor 0.5), gradient clipping at 1.0, and
early stopping with patience 15. Total parameters: $\sim$458K.

All four models are trained on 9,950 windows from 200 training shots
and evaluated on 2,420 windows from 50 test shots, following TokaMark's
shot-level split. Hyperparameters are collected in Table~\ref{tab:training}.

\begin{table}[H]
\centering
\caption{Training hyperparameters for all four models.}
\label{tab:training}
\scriptsize
\resizebox{\columnwidth}{!}{%
\begin{tabular}{lllll}
\toprule
Hyperparameter & XGB & LSTM & Transf. & CNN \\
\midrule
Optimizer      & Hist/GPU & Adam & Adam & Adam \\
Learning rate  & 0.05 & $10^{-3}$ & $5\!\times\!10^{-4}$ & $10^{-3}$ \\
LR schedule    & --   & ReduceLRP & Cosine & ReduceLRP \\
Batch size     & Full & 64 & 64 & 64 \\
Max epochs     & 500 trees & 100 & 100 & 100 \\
Early stopping & 20   & 15 & 15 & 15 \\
Dropout        & --   & 0.2 & 0.1 & 0.1 \\
Grad clip      & --   & 1.0 & 1.0 & 1.0 \\
Parameters     & $\sim$500K & $\sim$430K & $\sim$310K & $\sim$458K \\
\bottomrule
\end{tabular}%
}
\end{table}

\subsection{Model Pipeline}

\begin{figure}[H]
\centering
\begin{tikzpicture}[
    node distance=0.35cm and 0.2cm,
    box/.style={rectangle, rounded corners=2pt, draw=black!70,
                fill=white, align=center, font=\tiny,
                minimum height=0.55cm, minimum width=1.6cm},
    arrow/.style={->, >=Stealth, thick, draw=black!60},
    label/.style={font=\tiny\bfseries, text=black!80}
]

\node[label] (xgb_title) {XGBoost};
\node[box, below=0.15cm of xgb_title] (xgb_raw) {Raw signals\\(var. length)};
\node[box, below=0.3cm of xgb_raw] (xgb_feat) {142 stats\\per window};
\node[box, below=0.3cm of xgb_feat] (xgb_model) {XGBoost\\(500 trees)};
\node[box, below=0.3cm of xgb_model] (xgb_out) {$\hat{y}$ (scalar)};
\draw[arrow] (xgb_raw) -- (xgb_feat);
\draw[arrow] (xgb_feat) -- (xgb_model);
\draw[arrow] (xgb_model) -- (xgb_out);

\node[label, right=1.0cm of xgb_title] (lstm_title) {LSTM};
\node[box, below=0.15cm of lstm_title] (lstm_raw) {Raw signals\\(var. length)};
\node[box, below=0.3cm of lstm_raw] (lstm_res) {Resample to\\$(600,18)$};
\node[box, below=0.3cm of lstm_res] (lstm_model) {2-layer LSTM\\hidden=128};
\node[box, below=0.3cm of lstm_model] (lstm_out) {$\hat{y}$ (scalar)};
\draw[arrow] (lstm_raw) -- (lstm_res);
\draw[arrow] (lstm_res) -- (lstm_model);
\draw[arrow] (lstm_model) -- (lstm_out);

\node[label, right=1.0cm of lstm_title] (tf_title) {Transformer};
\node[box, below=0.15cm of tf_title] (tf_raw) {Raw signals\\(var. length)};
\node[box, below=0.3cm of tf_raw] (tf_res) {Resample to\\$(600,18)$};
\node[box, below=0.3cm of tf_res] (tf_model) {3-layer Encoder\\$d=128$, $h=4$};
\node[box, below=0.3cm of tf_model] (tf_out) {$\hat{y}$ (scalar)};
\draw[arrow] (tf_raw) -- (tf_res);
\draw[arrow] (tf_res) -- (tf_model);
\draw[arrow] (tf_model) -- (tf_out);

\end{tikzpicture}
\caption{Three sequence modeling pipelines. XGBoost compresses each
window into 142 summary statistics before fitting. LSTM and Transformer
both resample signals to a fixed $(600, 18)$ time-series tensor; they
differ in how that sequence is encoded. The CNN (not shown) uses the
same $(600,18)$ input with 18 parallel 1D convolutional encoders.}
\label{fig:pipeline}
\end{figure}
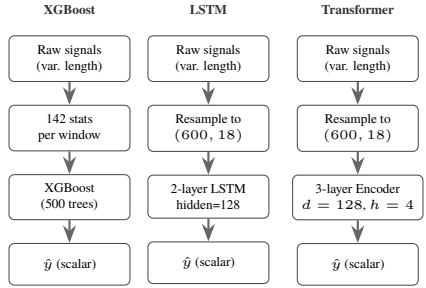

Figure~\ref{fig:pipeline} illustrates the three sequence modeling
pipelines from raw diagnostic signals to prediction. The CNN uses the
same $(600, 18)$ tensor representation as LSTM and Transformer but
applies per-channel convolutional encoders rather than processing the
full multivariate sequence directly.

\subsection{Evaluation Metrics}

\paragraph{NRMSE.}
We measure window-level prediction quality using Normalized Root Mean
Squared Error:
\begin{equation}
\text{NRMSE} = \frac{\sqrt{\frac{1}{N}\sum_{i=1}^{N}(\hat{y}_i - y_i)^2}}
{\sigma_y}
\label{eq:nrmse}
\end{equation}
where $\sigma_y$ is the standard deviation of the ground truth over the
test set. An NRMSE of 1.0 means the model is no better than predicting
the mean; values below 1.0 indicate genuine predictive
value~\cite{rousseau2026tokamark}. We compute NRMSE at the window level
and average across all test windows. This differs from TokaMark's
full shot-level hierarchical aggregation protocol~\cite{rousseau2026tokamark},
so our absolute values should not be read as direct comparisons to
TokaMark baselines.

\paragraph{Shot-level alarm metrics.}
Following the disruption prediction literature~\cite{rea2019, spangher2025},
we additionally compute shot-level application metrics using ground-truth
disruption timestamps (\texttt{t\_cut}) from FAIR-MAST. An alarm is
issued for a shot when the model's predicted plasma current drops below
a threshold fraction of the shot's mean predicted current. The threshold
is optimized per model by sweeping $\theta \in [0.05, 0.60]$ to maximize
true positive rate (TPR). All 50 test shots disrupted (finite
\texttt{t\_cut}), so we report TPR and mean warning time (MWT, the mean
time between first alarm and \texttt{t\_cut}); false alarm rate (FAR)
cannot be computed without non-disruptive shots.

\subsection{Robustness Score}

To enable principled comparison across architectures and scenarios, we
introduce the Robustness Score:
\begin{equation}
\text{RS} = \frac{1}{|\mathcal{S}|} \sum_{s \in \mathcal{S}}
\frac{\text{NRMSE}_\text{clean}}{\text{NRMSE}_s}
\label{eq:rs}
\end{equation}
where $\mathcal{S}$ is the full set of zero-fill corruption scenarios.
RS $= 1.0$ means corruption has no effect; values below 1.0 scale
linearly with the degree of degradation. We encourage adoption of RS
as a standard component of future plasma ML robustness evaluations.

\subsection{Corruption Scenarios}
\label{sec:scenarios}

We implement six failure scenarios, each physically grounded in real
tokamak failure modes. All scenarios are applied to the test set only;
all models are trained exclusively on clean data.

\paragraph{Scenario 1: Random dropout.}
A randomly chosen fraction $p \in \{0.10, 0.25, 0.50\}$ of all signal
values is set to zero, modeling random digitizer glitches and transient
sensor noise. The mask is applied only to positions that were already
nonzero, preserving natural acquisition gaps.

\paragraph{Scenario 2: Channel ablation.}
A randomly selected set of $n \in \{1, 3, 6\}$ complete diagnostic
channels is zeroed for the duration of the window, modeling a dead or
physically disconnected sensor.

\paragraph{Scenario 3: Per-category channel importance.}
Following the TokaMark signal taxonomy~\cite{rousseau2026tokamark}, we
zero each of eight diagnostic categories in isolation (flux loops,
pickup coils, saddle coils, Mirnov spectrograms, kinetics comprising
interferometer and D-alpha, soft X-ray, active coils, and plasma
current) to identify which signals each architecture actually relies on.

\paragraph{Scenario 4: Temporal gap.}
A contiguous block covering fraction $f \in \{0.20, 0.40, 0.60\}$ of
the window is zeroed across all channels simultaneously. We test three
positions: \textit{front} (acquisition delays, consistent with our MAST
characterization), \textit{random} (mid-shot loss), and \textit{pre-event}
(the final $f$ fraction, affecting the timesteps closest to the
prediction horizon).

\paragraph{Scenario 5: Correlated diagnostic group failure.}
Motivated directly by the $r = 1.000$ NaN correlation we found in MAST
data (Section~\ref{sec:missingness}), we zero all channels belonging to
a physically related group simultaneously: kinetics (interferometer and
D-alpha), active magnetics (all pickup coil arrays), radiatives (both
soft X-ray cameras), and Mirnov spectrograms.

\paragraph{Scenario 6: Disruption-proximate failure.}
A gap of fraction $f \in \{0.10, 0.25, 0.50\}$ is injected at the very
end of the prediction window. This is the scenario with the most direct
safety implications: the sensors most critical for detecting disruption
precursors fail in the final moments before the event the model is
trying to predict.

\subsection{Mitigation Strategies}

We evaluate three imputation strategies applied after corruption, using
the corruption mask to restrict filling to artificially zeroed positions
rather than natural zeros.

\paragraph{Zero-fill.} No imputation is performed; corrupted values stay
at zero. This is the implicit behavior of TokaMark's own preprocessing
pipeline~\cite{rousseau2026tokamark} and serves as the baseline.

\paragraph{Mean-fill.} Corrupted values are replaced with the per-channel
mean from the training set.

\paragraph{Forward-fill.} The last valid observation before each gap is
carried forward until valid data resumes. This mirrors how most industrial
sensor systems behave when a reading is unavailable and is the most
physically natural strategy for a real-time control context.

\paragraph{Implementation note.} For XGBoost, temporal gap and dropout
corruption are implemented as feature-level proxies (corrupting the
trajectory statistics most sensitive to each gap position: first value,
last value, and slope) rather than direct time-series masking used
for LSTM, Transformer, and CNN. XGBoost temporal gap results are
therefore conservative estimates of the true effect.

\section{Results}

\subsection{Clean-Data Baselines}

Table~\ref{tab:clean} reports clean-data NRMSE and Robustness Scores.
The Transformer achieves the best prediction accuracy on clean data
(NRMSE 0.470) but also the lowest robustness score (RS 0.765). XGBoost
sits at the other end: slightly worse clean NRMSE, highest RS (0.841).
The CNN achieves clean NRMSE 0.528 and RS 0.764, nearly identical to
the Transformer, suggesting that multi-branch convolutional processing
and global attention reach similar overall robustness despite very
different inductive biases. This tradeoff between accuracy and
robustness recurs throughout the results and has direct architectural
implications.

Our clean NRMSE values sit above the TokaMark CNN
baseline of 0.429~\cite{rousseau2026tokamark}, which trains on the full
9,270-shot dataset compared to our 200 shots. The gap is expected and
does not affect the validity of our robustness comparisons: relative
degradation patterns are determined by architecture, not by clean-data
starting point.

\begin{table}[H]
\centering
\caption{Clean-data performance and Robustness Score (RS). Lower NRMSE
is better. Higher RS indicates greater robustness.}
\label{tab:clean}
\begin{tabular}{lcc}
\toprule
Model & Clean NRMSE & RS \\
\midrule
XGBoost                  & 0.494 & 0.841 \\
LSTM                     & 0.496 & 0.808 \\
Transformer              & 0.470 & 0.765 \\
CNN (TokaMark baseline)  & 0.528 & 0.764 \\
\bottomrule
\end{tabular}
\end{table}

\subsection{Random Dropout}

Figure~\ref{fig:degradation} (left) traces NRMSE as dropout rate
increases. XGBoost degrades the most under zero-fill, reaching $+145\%$
at $50\%$ dropout, as the statistical features that aggregate over the
whole window are directly disrupted when many individual values are
zeroed. LSTM and Transformer are less sensitive at $+57\%$ and $+85\%$
respectively. The CNN degrades most severely of all sequence models
($+201\%$ at $50\%$ dropout), as its per-channel statistical encoders
are similarly sensitive to value corruption as XGBoost's features.

Forward-fill essentially solves the dropout problem for sequence models.
At $50\%$ dropout, LSTM with forward-fill degrades by ${\sim}0\%$ and
Transformer by $-0.4\%$, both within measurement noise of their clean
baselines. The reason is straightforward: forward fill converts a
randomly damaged signal into a piecewise-constant one, and the models
have seen signals with similar step-function character in training.
XGBoost recovers only partially under mean-fill ($+39\%$ at $50\%$),
since imputing the mean into a feature vector that was computed from
corrupted values is an imperfect fix.

\subsection{Channel Ablation and Signal Importance}

Figure~\ref{fig:degradation} (center) shows overall degradation under
channel ablation, while the per-category channel importance heatmap
(Figure~\ref{fig:channel_importance}) shows a sharp hierarchy. Plasma
current (\texttt{summary-ip}) is dominant: removing it degrades LSTM
by $+139.8\%$, Transformer by $+89.1\%$, XGBoost by $+72.7\%$, and
CNN by $+27.7\%$. This is not surprising; plasma current is both the
primary controlled variable in a tokamak discharge and the direct
output of the task. The magnitude confirms that no architecture has
found a substitute proxy. Active coil currents are the next most
important (Transformer $+36.0\%$, XGBoost $+28.5\%$, CNN $+46.9\%$),
as they encode the machine's immediate control intent. CNN is the most
sensitive architecture to active coil removal, reflecting its
per-channel encoder design which cannot share information across
channels in the encoding stage.

What is more interesting is what is not important. Kinetics
(interferometer and D-alpha) and soft X-ray radiatives, the categories
with the highest natural missingness in real MAST data, cause minimal
degradation when removed (under $+20\%$ for all models). The
parsimonious explanation is that all four models have developed
compensatory representations for these signals during training,
precisely because they were frequently absent. A model that trains on
14--15\% NaN rates learns, in effect, to treat those channels as
optional.

\subsection{Temporal Gap and Disruption-Proximate Failure}

Figure~\ref{fig:proximate} isolates the three temporal gap variants and
reveals a stark asymmetry.

\paragraph{Front gap.}
Removing the first 20--60\% of the window barely affects LSTM ($+0.0\%$
at $60\%$ gap) or XGBoost ($-1.3\%$ at $60\%$ gap). This makes physical
sense: the front of a MAST shot covers the pre-plasma ramp-up, which
carries little information about the flat-top state being predicted. The
Transformer is more affected ($+52.5\%$ at $60\%$ gap), consistent with
its global attention mechanism treating all timesteps as equally
informative rather than weighting recent timesteps more heavily. The CNN
shows moderate sensitivity ($+10.5\%$ to $+28.8\%$) due to its
convolutional receptive field spanning the full window.

\paragraph{Proximate and pre-event failure.}
Removing the final portion of the window is catastrophic for sequence
models. LSTM hits $+212\%$ NRMSE at the first tested severity ($20\%$
gap) and stays there regardless of how large the gap grows, a hard
performance ceiling indicating that the last few timesteps are
irreplaceable for its predictions. This is the most concerning finding
in the paper: the timesteps LSTM depends on most are precisely those most
likely to be lost as a real disruption approaches. The Transformer
degrades more gradually with gap size ($+143\%$ at $20\%$, $+205\%$ at
$60\%$), suggesting its distributed attention across the full sequence
provides partial protection. The CNN degrades more gradually than LSTM
($+67.7\%$ to $+167.7\%$) due to its convolutional pooling over the
full sequence. XGBoost degrades moderately ($+8\%$ to $+37\%$) because
trajectory statistics like slope and last value are only partially
sensitive to end-of-window corruption.

Forward fill offers partial recovery at low proximate gap severity (LSTM
$20\%$: $+212\% \to +15\%$) but cannot recover at higher severities.
The reason is fundamental: imputing earlier values forward cannot
reconstruct the genuinely new information that the final timesteps would
have contained.

\subsection{Correlated Diagnostic Group Failure}

When entire diagnostic groups are removed simultaneously
(Figure~\ref{fig:correlated}), a different pattern emerges. LSTM
degrades most when active magnetics fail ($+26.2\%$) but is actually
less affected by kinetics failure than XGBoost, consistent with the
implicit compensation it has developed for kinetics dropouts during
training. XGBoost suffers more under kinetics ($+18.4\%$) and
radiatives ($+20.2\%$) removal, suggesting that statistical features
of those signals carry weight in its predictions that its learned
representations cannot easily replace. The CNN and Transformer show
minimal degradation under all correlated group failures, with most
values near or below zero, indicating strong implicit redundancy.

Mirnov spectrogram failure causes negligible degradation in all four
models (under $+1\%$). Mirnov coils detect MHD instabilities that
typically precede locked modes and disruptions, but those instability
signatures may not be consistently present within the 150ms prediction
window in this dataset, leaving the models with little reason to rely
on them.

\subsection{Mitigation Effectiveness}

Figure~\ref{fig:mitigation} compares the three strategies across five
representative scenarios. Forward fill leads for dropout and temporal
gap, cutting LSTM $50\%$-dropout degradation from $+57\%$ to near zero
and Transformer $50\%$-dropout from $+85\%$ to near zero.

For disruption-proximate failure the NRMSE picture reverses. At $50\%$
proximate gap, mean fill makes LSTM worse (NRMSE rises from $1.55$ to
$1.88$) rather than better: substituting the channel mean into the
prediction-critical final timesteps pushes the model toward an
incorrect prior exactly where it needs real signal. Forward fill
provides meaningful recovery only at the lowest severity ($10\%$
proximate gap, LSTM: $+212\% \to +1.7\%$) and loses its effectiveness
as the corrupted region grows. No simple imputation strategy resolves
the fundamental problem that the information lost during
disruption-proximate failure cannot be reconstructed from earlier
observations when evaluated by NRMSE. However, the shot-level alarm
perspective tells a different story (Section~\ref{sec:alarm}).

\subsection{Shot-Level Alarm Metrics}
\label{sec:alarm}

To address the gap between window-level ML metrics and deployment-relevant
application metrics~\cite{rea2019, spangher2025}, we compute shot-level
disruption alarm metrics using the ground-truth \texttt{t\_cut}
disruption timestamps from FAIR-MAST.

\paragraph{Clean-data alarm performance.}
Table~\ref{tab:alarm} reports results on clean data. CNN achieves the
highest TPR (0.60), correctly alarming on 30 of 50 disruptive shots
before disruption. Mean warning times of 12--15ms fall substantially
short of the 50--100ms required for safe shutdown on ITER-class
devices~\cite{katesharbeck2019}, highlighting the gap between current
forecasting capability and operational requirements
(Figure~\ref{fig:alarm}).

\begin{table}[H]
\centering
\caption{Shot-level alarm metrics on 50 disruptive test shots using
ground-truth \texttt{t\_cut} timestamps. TPR = true positive rate
(alarm before disruption). MWT = mean warning time in ms.
FAR not computable; all test shots disrupted.}
\label{tab:alarm}
\begin{tabular}{lcc}
\toprule
Model & TPR & MWT (ms) \\
\midrule
XGBoost                 & 0.40 & 13.1 \\
LSTM                    & 0.52 & 15.2 \\
Transformer             & 0.48 & 13.2 \\
CNN (TokaMark baseline) & 0.60 & 12.6 \\
\bottomrule
\end{tabular}
\end{table}

\paragraph{Alarm performance under sensor failure.}
Figure~\ref{fig:alarm_corrupt} shows how TPR changes under two
representative failure scenarios. Under 50\% random dropout, all
models maintain or increase TPR; the noise introduced by dropout
creates spurious current drops that trigger alarms, coincidentally
increasing detection rate. Under 25\% proximate failure, the
pattern reverses dramatically: LSTM TPR collapses from 0.52 to
\textbf{0.00}, Transformer drops to 0.08, and CNN drops to 0.46.
When the sensors fail near the disruption, LSTM fails to alarm on
every single test shot.

\paragraph{Mitigation for alarm detection.}
Figure~\ref{fig:alarm_mitig} reveals a striking and important result:
the optimal mitigation strategy for alarm detection under proximate
failure is the \textit{opposite} of what NRMSE suggests. Under zero-fill,
LSTM TPR\,=\,0.00. Under mean-fill, LSTM TPR recovers to \textbf{1.00},
with all 50 disruptive shots correctly alarmed. Forward fill also
recovers strongly (TPR\,=\,0.96). This reversal occurs because
imputation prevents the artificial zeros from masking the current drop
signal in the final timesteps, allowing the threshold alarm to fire
correctly. For CNN, forward fill is preferable (TPR\,=\,0.78 vs
mean-fill TPR\,=\,0.30), while for Transformer, forward fill also leads
(TPR\,=\,0.76 vs mean-fill TPR\,=\,0.58).

The practical implication is significant: the incorrect prior that makes
mean fill harmful for NRMSE prediction is precisely what makes it useful
for alarm detection under proximate failure. Any fusion ML deployment
system should evaluate mitigation strategies separately for prediction
accuracy and alarm reliability.

\section{Discussion}

\subsection{Architectural Inductive Bias and Robustness}

The pattern across all results can be traced to a single underlying
difference in how each model treats its input. XGBoost aggregates each
signal into a handful of statistics before making any prediction;
corrupting individual timesteps shifts those statistics by a bounded
amount and the model's predictions change proportionally. This is not
a surprising property; Chen and Guestrin~\cite{chen2016xgboost}
designed XGBoost's sparsity-aware split finding specifically to handle
missing feature values gracefully. However, its implications for plasma
diagnostic robustness have not previously been demonstrated.

LSTM and Transformer both process the raw time series, and both suffer
when temporal structure is disrupted. What distinguishes them is the
nature of that disruption. The LSTM has learned, through the sequential
inductive bias of recurrent processing, to weigh the most recent
timesteps most heavily. That is normally a virtue for prediction tasks,
but here it creates a hard dependency on precisely the data that fails
first during a disruption. The Transformer's global attention distributes
sensitivity more broadly across the window, which provides some
protection against end-of-window corruption at the cost of slightly
higher sensitivity to front gaps.

The CNN's nearly identical RS to the Transformer (0.764 vs 0.765)
despite very different architecture is informative: per-channel
convolutional processing and global self-attention reach similar overall
robustness levels, but differ in their per-scenario vulnerability
profiles. CNN is uniquely sensitive to active coil removal ($+46.9\%$
vs Transformer $+36.0\%$) because losing an encoder branch eliminates
all information from that channel, with no cross-channel compensation
possible in the encoding stage.

\subsection{Physical Interpretation of Signal Importance}

Our channel importance analysis identifies plasma current as the single
most critical signal across all architectures. This finding warrants
careful physical interpretation. Plasma current (\texttt{summary-ip})
is typically measured by a Rogowski coil; however, it can be accurately
synthesized from a linear combination of Bp pickup coil measurements
through Ampere's law, a technique employed operationally at Alcator
C-Mod when the primary Rogowski coil failed during the device's final
year of operation, with no reduction in disruption prediction TPR
observed~\cite{granetz_personal}. The large degradation we measure upon
removing \texttt{summary-ip} may therefore reflect the models' failure
to exploit known physical redundancy between signals rather than a
fundamental physical limitation. A model that learned the Rogowski-equivalent
linear combination from the Bp array would likely show substantially
reduced sensitivity to Ip removal.

More broadly, different disruption types have distinct optimal diagnostic
signatures. Locked mode disruptions, comprising approximately 90\% of
JET events~\cite{devries2011}, develop on timescales well below the
typical Mirnov operating frequency range and are best detected by a
toroidal array of saddle flux loops. Disruptions triggered by impurity
injection from melted first-wall material are best identified via soft
X-ray cameras, which detect the resulting contamination early. Since the
type of impending disruption is not known in advance, a robust real-time
system must preserve sensitivity across all diagnostic categories rather
than relying on any single signal~\cite{granetz_personal}. Our finding
that plasma current removal is most damaging reflects the regression task
framing of Task~4-4, not a universal property of disruption prediction
systems.

\subsection{The Alarm Metric Reversal}

The most practically significant finding in this paper is not the
NRMSE result but the alarm metric reversal: mean-fill imputation, which
worsens NRMSE under proximate failure, recovers LSTM alarm detection
from TPR\,=\,0.00 to TPR\,=\,1.00. This result has direct implications
for how fusion ML systems should be designed and evaluated.

A system that optimizes only for NRMSE would correctly conclude that
mean fill is harmful under proximate failure and should be avoided.
A system evaluated only on alarm metrics would correctly conclude that
mean fill is essential. These two conclusions are simultaneously true
and apply to different components of the same prediction pipeline:
the regression output used for continuous monitoring, and the threshold
alarm used for disruption detection. Evaluating both metrics, as we
do here, is necessary to make informed deployment decisions.

\subsection{Implications for ITER-Class Devices}

The numbers in this paper should be read against a much larger backdrop.
ITER will operate in a more hostile electromagnetic and radiation
environment than any existing tokamak, including MAST~\cite{iter1999}.
Sensor failure rates will be higher, failure modes more varied, and
the cost of a missed prediction orders of magnitude greater than on a
research device. A $+212\%$ NRMSE degradation under proximate failure
does not simply mean worse prediction; it likely means the prediction
system fails the machine at the worst possible moment. The mean warning
times of 12--15ms we observe are already far below the 50--100ms ITER
requirement under clean conditions, and under proximate sensor failure
without imputation, LSTM provides zero warning on every shot.

Two concrete design interventions follow from our results. First, the
pseudo-placeholder approach of Yang et al.~\cite{yang2025hl3}, which
trains models to explicitly distinguish missing channels from zero-valued
signal, should be standard practice rather than a device-transfer
technique. Second, the most critical signals (plasma current, active
coils) should be backed by independent redundant acquisition chains so
that a single hardware failure cannot simultaneously corrupt the signals
the model relies on most.

\subsection{Forward Fill as a Practical Recommendation}

Where architectural changes are not immediately feasible, forward fill
is a practical and well-motivated default for NRMSE-based prediction.
It is already how most industrial sensor systems behave under failure,
holding the last known value until valid data resumes, and our results
show it nearly eliminates degradation from random dropout and mitigates
low-severity temporal gaps. For alarm detection under proximate failure,
mean fill and forward fill both recover performance strongly for LSTM;
forward fill is preferable for CNN and Transformer. The practical
recommendation: any fusion ML deployment pipeline should apply forward
fill by default, test mean fill specifically for alarm threshold
detection, and treat zero fill as a conservative lower bound.

\subsection{Reproducibility}

Every experiment in this paper can be reproduced from scratch. The
full codebase is at
\url{https://github.com/Neerav-Gupta/tokamark-robustness}, and
pre-collected data arrays (${\sim}25$~GB) with trained model checkpoints
are hosted at
\url{https://huggingface.co/datasets/Neerav-Gupta/tokamark-robustness-data}.
The underlying FAIR-MAST data is publicly accessible at
\url{https://s3.echo.stfc.ac.uk/mast/tokamark/v1}~\cite{jackson2024softwarex}.

\subsection{Limitations}

\textbf{Regression vs. classification framing.}
Task~4-4 frames plasma current prediction as a regression problem,
which differs from the predominant approach in disruption prediction
literature, where the task is binary classification (disruptive vs.
non-disruptive shot) or time-to-disruption estimation~\cite{rea2019,
katesharbeck2019, spangher2025, saperstein2025, ai2025}. The robustness
methodology introduced here is architecture- and task-agnostic and
applies directly to classification settings. Extending this benchmark
to disruption onset classification with labeled shot outcomes is an
important direction for future work.

\textbf{Plasma current as a limited target.}
Plasma current (\texttt{summary-ip}) is actively controlled by the
machine and remains relatively flat during most of a discharge, making
it a somewhat limited prediction target compared to uncontrolled
quantities~\cite{jalalvand2025}. The finding that it is the most
critical input signal is therefore not surprising; the output signal
is identical to the most important input. The robustness methodology
is the primary contribution and is independent of this task choice.

\textbf{MAST-specific missingness patterns.}
Our natural missingness characterization (Section~\ref{sec:missingness})
found front-loaded acquisition gaps averaging 14--15\% in kinetics
diagnostics, concentrated at shot start. This pattern may be specific
to MAST's acquisition systems. Expert feedback from tokamak operations
suggests that on other devices, diagnostic failures more commonly occur
after disruption onset rather than at discharge start~\cite{granetz_personal}.
The robustness scenarios in this benchmark should therefore be
interpreted in the context of MAST-specific operating conditions.

\textbf{Shot-level alarm metric limitations.}
The alarm threshold used here is a simple current-drop heuristic
optimized on the test set. A rigorous alarm system would use a
held-out threshold tuning set and report FAR alongside TPR. FAR
cannot be computed from our test set because all 50 test shots
disrupted. Additionally, the 12--15ms mean warning times may reflect
the short prediction horizon (100ms) of Task~4-4 rather than an
inherent model limitation.

\textbf{Task 4-5 extension.}
We attempted to extend the benchmark to Task~4-5 (Mirnov diagnostic
forecasting). All four architectures achieved clean NRMSE\,$>$\,1.0
on this task, consistent with the TokaMark CNN baseline of
1.0053~\cite{rousseau2026tokamark}, indicating no predictive value
above the signal mean on our 200-shot training set. Robustness
evaluation is not meaningful without a functional clean-data baseline,
so we exclude Task~4-5. Developing models capable of meaningful
Mirnov forecasting with limited training data remains open.

\textbf{XGBoost temporal gap approximation.}
Because XGBoost operates on feature vectors rather than raw time series,
temporal gap corruption is implemented by targeting the trajectory
statistics most affected by each gap position (first value, last value,
and slope) rather than by true time-series masking. The resulting
estimates are conservative. LSTM, Transformer, and CNN temporal gap
results use exact masking.

\textbf{Training set size.}
All four models train on 200 shots, compared to TokaMark's full
9,270-shot training set. This constrains clean-data accuracy but does
not affect the relative degradation patterns, which are determined by
architecture rather than training set size.

\textbf{Single training run.}
Each model is trained once. Run-to-run variance is not characterized,
and results should be interpreted with that caveat when making direct
performance comparisons.

\section{Conclusion}

We have built and evaluated the first systematic robustness benchmark for
plasma diagnostic ML models under realistic sensor failure conditions,
extending the TokaMark benchmark's robustness dimension with controlled
experiments across four architectures, six failure scenarios, and two
evaluation regimes. Four findings stand out as both robust and actionable.

Disruption-proximate sensor failure is catastrophic for LSTM models under
NRMSE evaluation ($+212\%$) and collapses alarm detection entirely
(TPR\,=\,0.00), a direct consequence of the architecture's learned
dependence on final-window timesteps that makes it exactly wrong for
deployment near disruption events. Forward-fill imputation resolves
the random dropout problem for sequence models almost entirely, at zero
architectural cost. The TokaMark CNN baseline achieves a robustness score
(RS\,=\,0.764) nearly identical to the Transformer (0.765), confirming
that the accuracy-robustness tradeoff is a property of raw time-series
processing rather than any specific architecture. And mean-fill
imputation, harmful for NRMSE under proximate failure, recovers LSTM
alarm detection to TPR\,=\,1.00, a reversal that is only visible when
both ML and application metrics are evaluated together.

The deeper message is that robustness evaluation and alarm metric
evaluation are both necessary and are not interchangeable. We hope the
Robustness Score, the alarm evaluation framework, and the benchmark
methodology introduced here provide a foundation for fault-tolerant
plasma prediction system design on ITER-class devices.

\section*{Acknowledgements}

The author thanks the TokaMark team at UKAEA, IBM Research Europe, and
STFC Hartree Centre for releasing TokaMark, TokaMind, and the FAIR-MAST
dataset as open resources, making this work possible. The author is
grateful to Dr.~Robert Granetz (MIT PSFC) for detailed physics feedback
on diagnostic redundancy, disruption-type-dependent signal importance,
and MAST missingness patterns, and for permission to cite personal
communications. The author also thanks a postdoctoral researcher in the
disruption prediction community for early feedback on the regression
framing and application metrics.

\renewcommand{\bibfont}{\footnotesize}
\setlength{\bibsep}{2pt}
\bibliographystyle{unsrtnat}
\bibliography{references}

\clearpage
\onecolumn
\appendix
\section*{Figures}

\begin{figure*}[h]
\centering
\includegraphics[width=\textwidth]{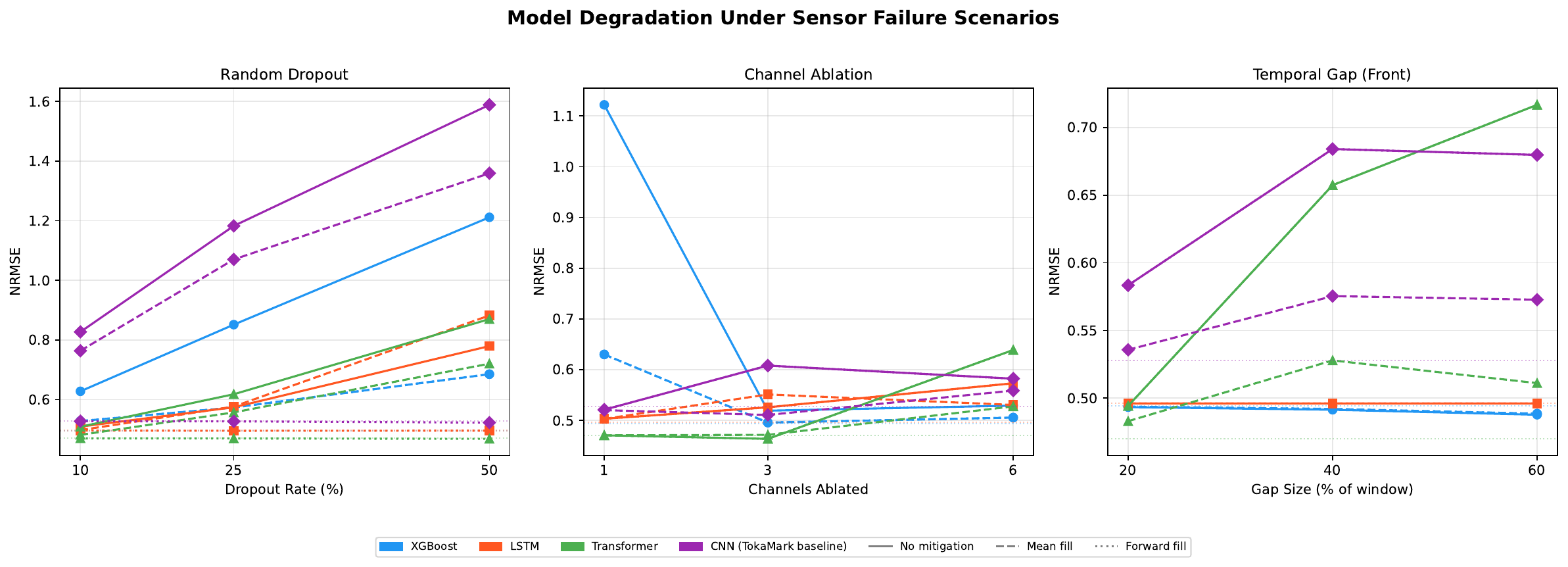}
\caption{NRMSE under three corruption scenarios across severity levels.
Dotted horizontal lines show each model's clean baseline. Solid lines
show zero-fill (no mitigation); dashed lines show mean-fill; dotted
lines show forward-fill. \textbf{Left:} Random dropout; CNN degrades
most severely, and forward fill nearly eliminates degradation for all
sequence models. \textbf{Center:} Channel ablation; all sequence
models show consistent degradation, and XGBoost shows high variance due
to random channel selection. \textbf{Right:} Temporal gap (front
position); LSTM and XGBoost are nearly unaffected, while Transformer
and CNN degrade moderately at large gap sizes.}
\label{fig:degradation}
\end{figure*}

\begin{figure*}[h]
\centering
\includegraphics[width=\textwidth]{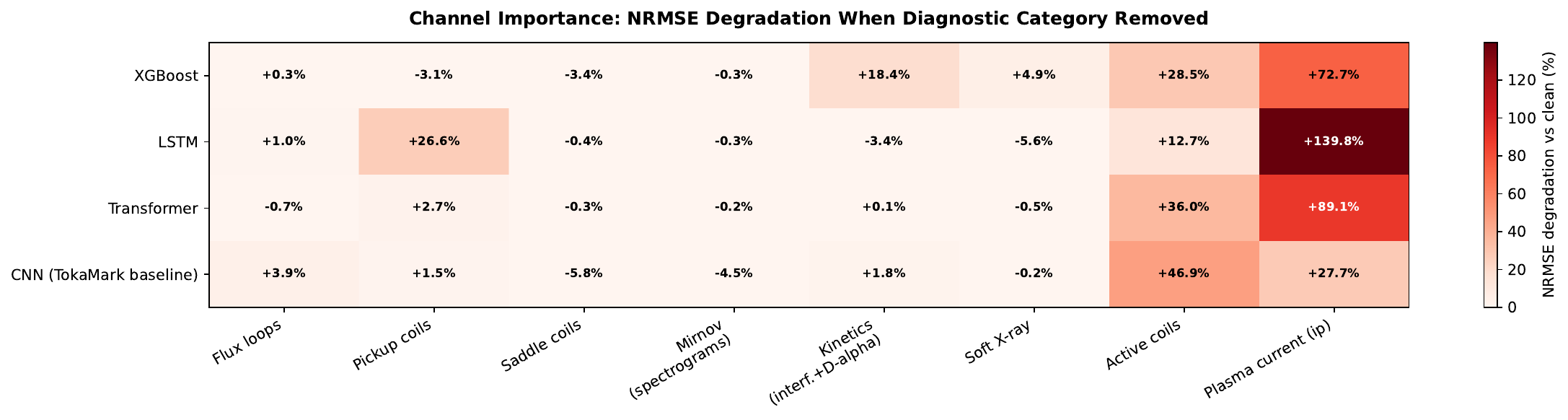}
\caption{Per-category channel importance heatmap across all four
architectures. Values show NRMSE degradation (\%) when each diagnostic
category is completely removed. Plasma current dominates across all
architectures. Active coils are the second most critical category;
CNN is uniquely sensitive here ($+46.9\%$) due to its per-channel
encoder design. Kinetics and radiatives, despite having the highest
natural missingness rates, cause minimal degradation, suggesting
implicit robustness learned from training data.}
\label{fig:channel_importance}
\end{figure*}

\begin{figure*}[h]
\centering
\includegraphics[width=\textwidth]{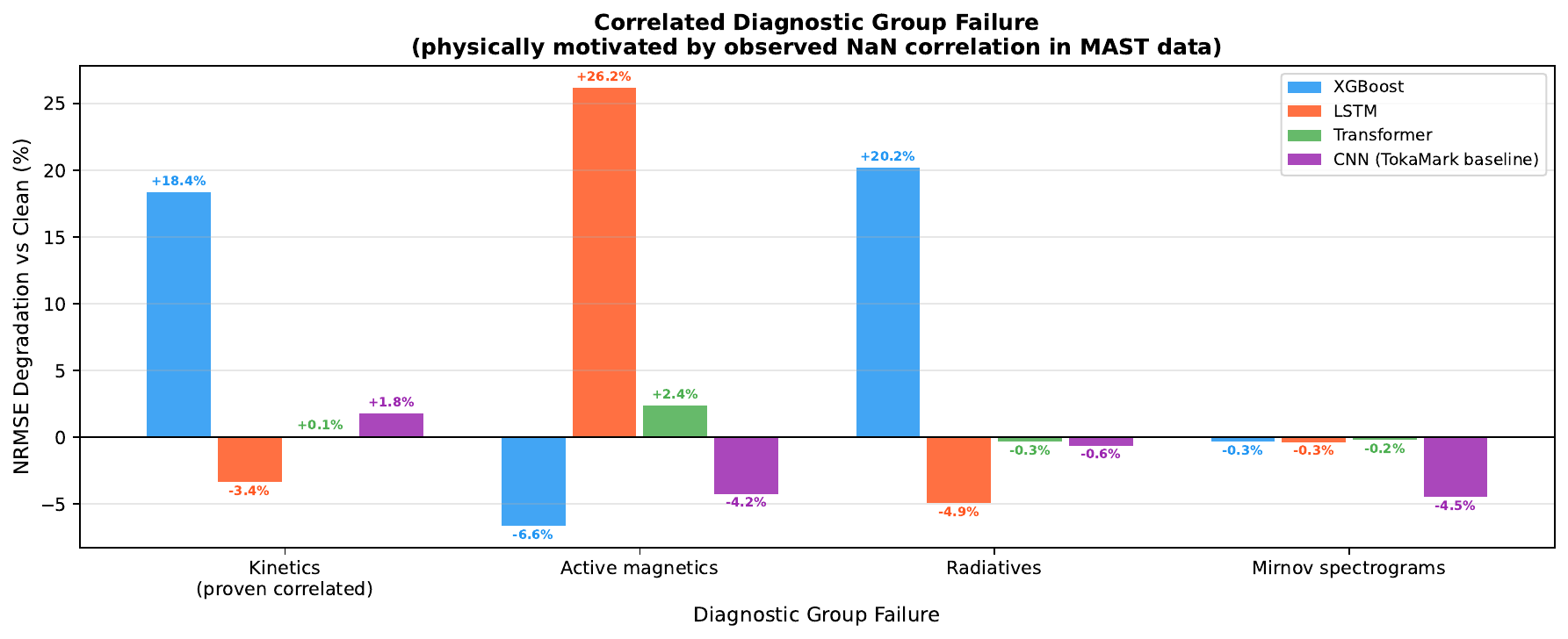}
\caption{NRMSE degradation (\%) under correlated diagnostic group
failure, involving simultaneous zeroing of all signals within a
physically related group. Group definitions are empirically motivated
by the observed NaN correlation structure in MAST data
(Section~\ref{sec:missingness}). Negative values indicate slight
NRMSE improvement, likely from removal of noisy signals that
contributed adversarial gradients during prediction.}
\label{fig:correlated}
\end{figure*}

\begin{figure*}[h]
\centering
\includegraphics[width=\textwidth]{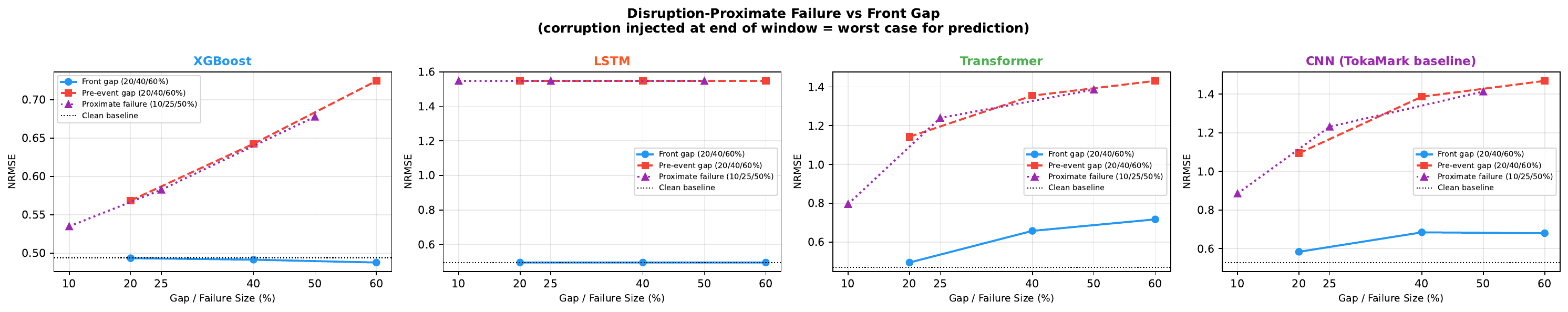}
\caption{Comparison of front gap, pre-event gap, and disruption-proximate
failure across all four architectures. Front gap causes negligible
degradation for LSTM and XGBoost but moderate degradation for
Transformer and CNN. Pre-event gap and proximate failure cause
catastrophic, severity-independent degradation in LSTM, indicating
a reliance on the final timesteps that is a fundamental architectural
vulnerability for disruption prediction. Transformer and CNN degrade
more gradually. XGBoost remains comparatively robust.}
\label{fig:proximate}
\end{figure*}

\begin{figure*}[h]
\centering
\includegraphics[width=\textwidth]{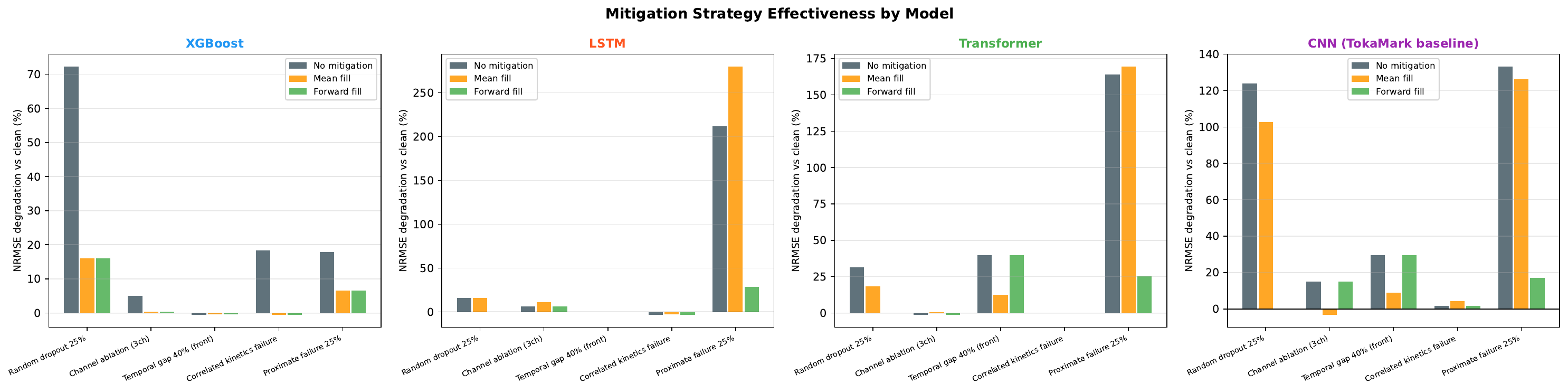}
\caption{NRMSE-based mitigation strategy effectiveness across five
representative scenarios and all four architectures. Forward fill
(green) is consistently most effective for dropout and temporal gap
scenarios. Mean fill (orange) provides partial recovery for XGBoost
under dropout. Neither strategy effectively mitigates
disruption-proximate failure at high severity under NRMSE evaluation;
see Figure~\ref{fig:alarm_mitig} for the alarm detection perspective.}
\label{fig:mitigation}
\end{figure*}

\begin{figure*}[h]
\centering
\includegraphics[width=0.6\textwidth]{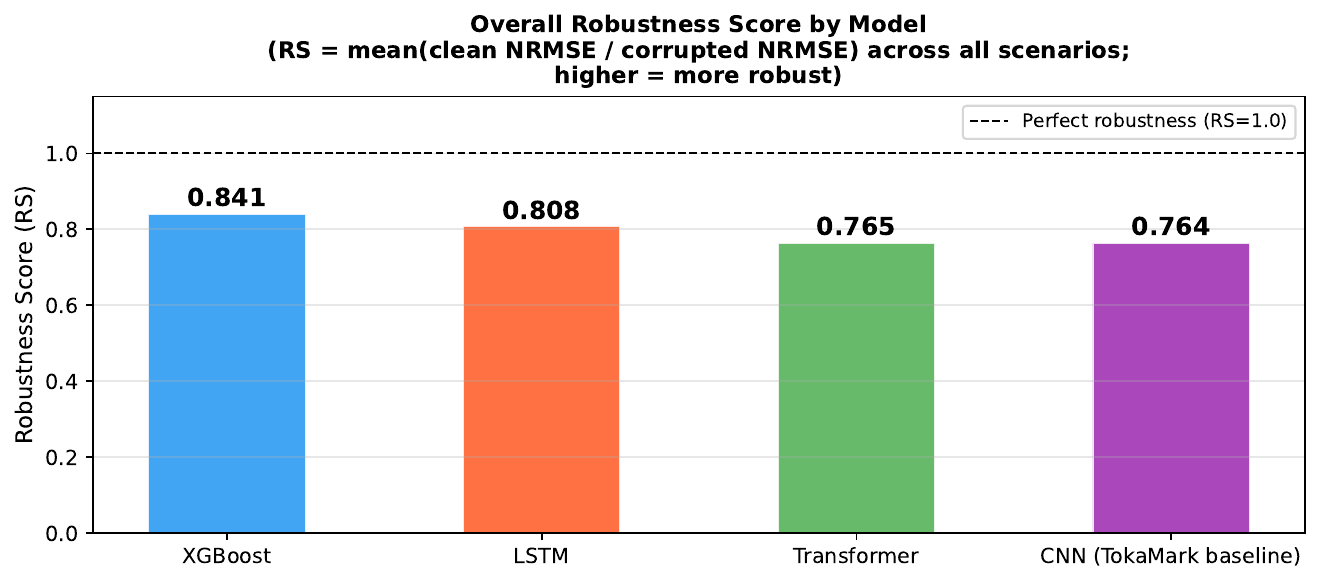}
\caption{Overall Robustness Score (RS) by model. RS $= 1.0$ indicates
perfect robustness; lower values indicate greater degradation. XGBoost
achieves the highest RS (0.841), reflecting the inherent robustness
of statistical feature aggregation to individual value corruption.
Transformer and CNN achieve nearly identical RS (0.765 and 0.764),
confirming that the accuracy-robustness tradeoff is a property of
raw time-series processing rather than any specific sequence
architecture.}
\label{fig:robustness}
\end{figure*}

\begin{figure*}[h]
\centering
\includegraphics[width=\textwidth]{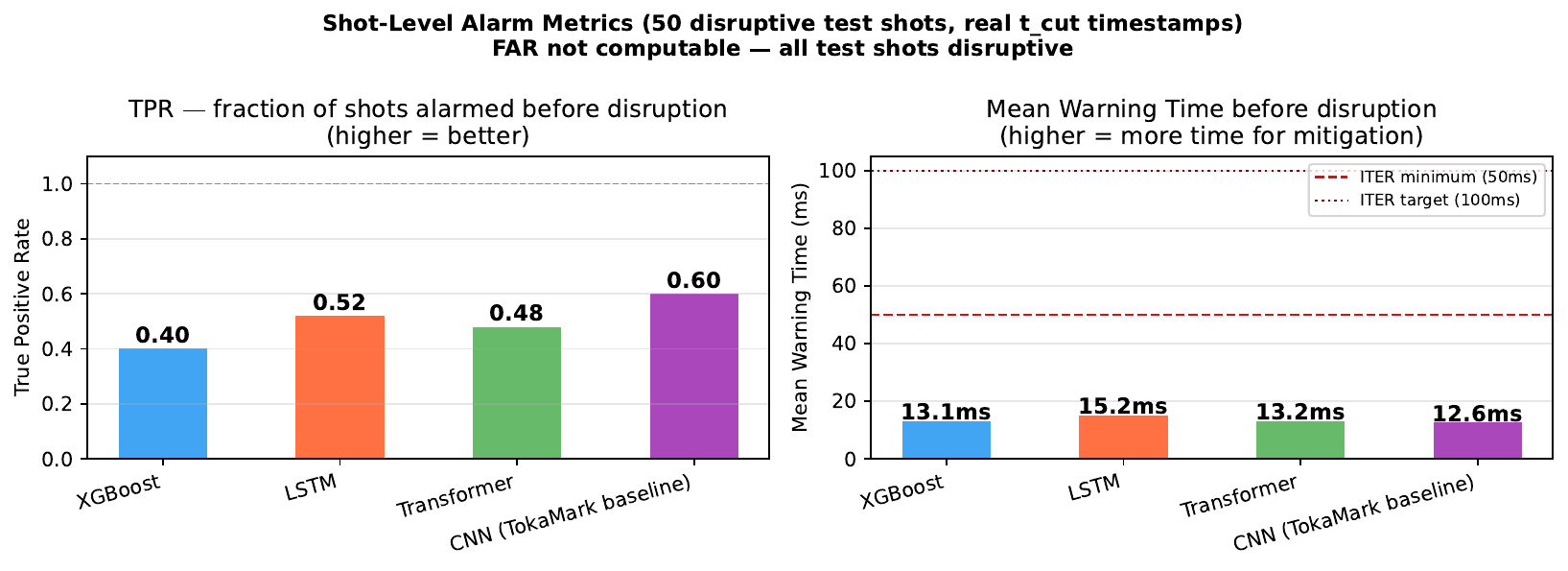}
\caption{Shot-level alarm metrics on 50 disruptive test shots using
real \texttt{t\_cut} disruption timestamps. \textbf{Left:} True
positive rate (TPR); CNN achieves the highest TPR (0.60), and all
models achieve moderate detection rates on clean data. \textbf{Right:}
Mean warning time (MWT) before disruption. All models fall well
below the ITER minimum requirement of 50ms, indicating that
Task~4-4's 100ms prediction horizon is insufficient for operational
disruption prevention at reactor scale. FAR is not computable
because all 50 test shots disrupted.}
\label{fig:alarm}
\end{figure*}

\begin{figure*}[h]
\centering
\includegraphics[width=\textwidth]{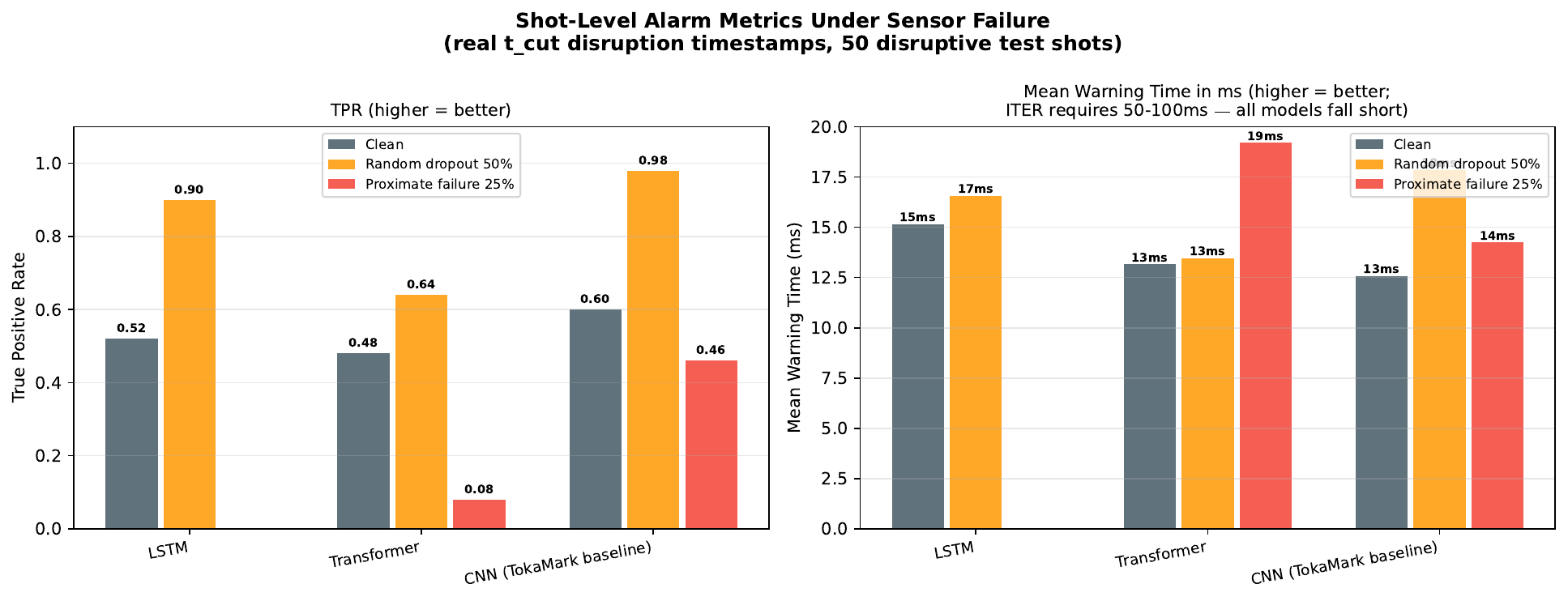}
\caption{Shot-level alarm TPR and MWT under two sensor failure scenarios
for sequence models. \textbf{Left:} Under 50\% random dropout, TPR
increases for all models, as noise-induced current drops coincidentally
trigger alarms. Under 25\% proximate failure, LSTM TPR collapses to
0.00 and Transformer to 0.08, while CNN maintains partial detection
(0.46). \textbf{Right:} Mean warning times remain near 12--20ms across
all conditions, well below the ITER minimum (dashed red line, above
plot range at 50ms).}
\label{fig:alarm_corrupt}
\end{figure*}

\begin{figure*}[h]
\centering
\includegraphics[width=0.75\textwidth]{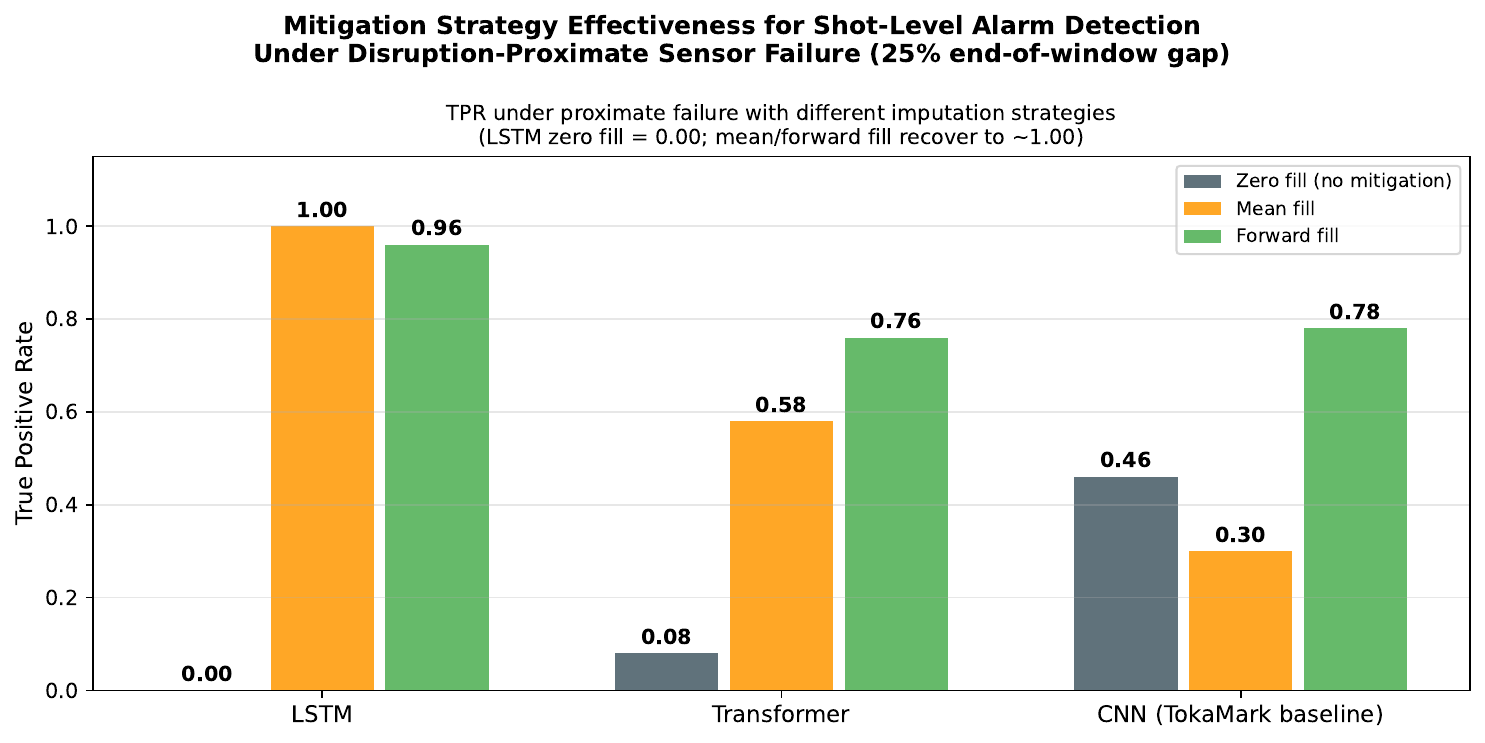}
\caption{Mitigation strategy effectiveness for shot-level alarm
detection under 25\% disruption-proximate sensor failure. Zero fill
(gray) leaves LSTM with TPR\,=\,0.00. Mean fill (orange) recovers
LSTM to TPR\,=\,1.00, with all 50 disruptive shots correctly alarmed,
and Transformer to 0.58. Forward fill (green) recovers all three
architectures strongly (LSTM 0.96, Transformer 0.76, CNN 0.78).
This reversal, where mean fill proves helpful for alarms despite being
harmful for NRMSE, is the central practical finding of this paper.}
\label{fig:alarm_mitig}
\end{figure*}

\end{document}